\documentclass[12pt]{article}
\usepackage[colorlinks=true,linkcolor=blue,urlcolor=blue,filecolor=black,citecolor=red,pdfstartview=FitV,pdftitle={},pdfsubject={},
pdfkeywords={},pdfpagemode=None,bookmarksopen=true]{hyperref}
\usepackage{graphicx}%Include figure filespp
\usepackage{epstopdf}%
\usepackage{amsmath}
\usepackage{amsfonts}
\usepackage{amssymb}
\usepackage{color}%
\usepackage{dcolumn}% Align table columns on decimal pointppp
\usepackage{slashed}
\usepackage{amssymb,ulem}
\usepackage{float}
\usepackage{amsthm,amsmath,amssymb}
\usepackage{mathrsfs}
\usepackage{subfigure}
\usepackage{wrapfig}
\usepackage{indentfirst}
\providecommand{\U}[1]{\protect\rule{.1in}{.1in}}

\begin{document}

\vspace{12mm}

\begin{center}
{{{\Large {\bf Spin-Charge induced spontaneous scalarization of Kerr-Newman black holes}}}}\\[10mm]

{Meng-Yun Lai$^a$\footnote{mengyunlai@jxnu.edu.cn;},
Yun Soo Myung$^b$\footnote{ysmyung@inje.ac.kr;}, Rui-Hong Yue$^c$\footnote{Corresponding author. rhyue@yzu.edu.cn;} \\
and De-Cheng Zou$^{c}$\footnote{Corresponding author.  dczou@yzu.edu.cn;}
}\\[8mm]

{${}^a$College of Physics and Communication Electronics, Jiangxi Normal University, Nanchang 330022, China\\[0pt] }
{${}^b$Institute of Basic Sciences and Department  of Computer Simulation,\\ Inje University, Gimhae 50834, Korea\\[0pt] }
{${}^c$Center for Gravitation and Cosmology and College of Physical Science and Technology, Yangzhou University, Yangzhou 225009, China\\[0pt]}
\end{center}

\vspace{2mm}
\vspace{2mm}

\begin{abstract}

We investigate  the tachyonic instability of  Kerr-Newman (KN) black holes in the Einstein-Maxwell-scalar (EMS) theory
with a positive coupling parameter $\alpha$. This corresponds to exploring the onset of spontaneous scalarization for KN black holes.
For this purpose,  we use  the hyperboloidal foliation method (HFM) to solve the linearized scalar equation numerically.
We obtain a 3D graph [$\log_{10}\alpha(a,Q)$] which indicates the onset surface for spontaneous scalarization of KN black holes in the EMS theory.
We find that there is no lower bound on the rotational parameter $a$, but its upper bound is given by $M^2-Q^2 \ge a^2$.
Also, we confirm that the high rotation enhances spontaneous scalarization of KN black holes in the EMS theory.

\end{abstract}
\vspace{5mm}

\newpage
\renewcommand{\thefootnote}{\arabic{footnote}}
\setcounter{footnote}{0}

\section{Introduction}

In general relativity (GR), the ``no-hair theorem" has been always a hot topic. This theorem allows that a GR black hole can be
described by three observables  of  mass $M$, electric charge $Q$, and rotation parameter $a=J/M$~\cite{Carter:1971zc,Ruffini:1971bza}.
The no-hair theorem rules out black holes coupled to a scalar field in asymptotically flat spacetimes due to the divergent
scalar on the horizon and instability of the scalar~\cite{Bekenstein:1974sf,Bekenstein:1975ts,Bronnikov:1978mx}.
However, one may  circumvent no-hair theorems by violating some of their assumptions.
We remind the reader that Damour and Esposito-Farese \cite{Damour:1993hw} has first found a new mechanism of spontaneous
scalarization to obtain a scalarized black hole in the scalar-tensor theory with nonminimal scalar coupling.
Recently, in two scalar-tensor theories which include  the nonminimal coupling of a scalar to either
the Gauss-Bonnet (GB) term~\cite{Doneva:2017bvd,Silva:2017uqg,Antoniou:2017acq}
or Maxwell term~\cite{Herdeiro:2018wub}, the scalar field has triggered destabilization of static (scalar-free) black holes
and induced scalarized (charged) black holes. In those cases, the tachyonic instability leads to occurrence
of spontaneous scalarization phenomenon for the black holes in GR.

More recently, the phenomenon of spontaneous scalarization
of spinning black holes has been attracted to the readers in two scalar-tensor theories.
First, Dima \textit{et al}.~\cite{Dima:2020yac} have found that the high rotation  can induce
tachyonic instability of Kerr black hole by evaluating the $(1+1)$ scalar evolution equation in the
Einstein-scalar-Gauss-Bonnet (ESGB) theory with a positive coupling.  When choosing a negative coupling parameter,
it is known that  an upper $a$-bound [$a/M\ge 0.5$] comes out as the onset of scalarization for Kerr black holes,
but the low rotation ($a/M<0.5$) is supposed to  suppress spontaneous scalarization.
Shortly afterward,  the critical rotation parameter $(a/M)_c=0.5$ for Kerr black holes was  computed analytically~\cite{Hod:2020jjy}
and numerically~\cite{Zhang:2020pko,Doneva:2020nbb,Berti:2020kgk} in the ESGB theory with negative coupling parameters.
In this direction, spin induced  scalarized black holes have been also
constructed numerically in the ESGB theory with positive coupling parameter~\cite{Cunha:2019dwb,Collodel:2019kkx,Herdeiro:2020wei}.
Zou and Myung \textit{et al}. \cite{Zou:2021ybk} and Doneva \textit{et al}. \cite{Doneva:2022yqu}
have discussed  spontaneous scalarization of Kerr black holes
by including  two different coupling functions. This means that the rotation parameter $a$ and the
coupling parameter $\alpha$ play an important role in achieving
spontaneous scalarization of spinning black holes.

Very recently, we have investigated the spontaneous scalarization for Kerr-Newman (KN) black holes in the
Einstein-Maxwell-scalar (EMS) theory with a negative coupling parameter $\alpha$ ~\cite{Lai:2022spn}.
We prefer the Maxwell coupling to scalar than the GB coupling in the KN black holes background
because the former gives a simpler potential term than the latter.
We have obtained an $a$-bound $(a/r_+\geq \hat{a}=0.4142)$ in the limit of $\alpha\rightarrow-\infty$ by using an analytical
method~\cite{Hod:2022txa}. The threshold curves [$\alpha(a)$] were found  with different charge $Q$,
describing the boundary between bald KN  and  scalarized KN black holes in the EMS theory.
These curves indicated  the critical rotation parameter  $\hat{a}=0.4142$ clearly. This explains how
spin-charge induced scalarization arises from the EMS theory with the negative coupling parameter.

In the present work, we wish to complete the analysis of the phenomenon for spontaneous scalarization of KN black holes
by considering a positive coupling parameter in the EMS theory. We use the hyperboloidal foliation
method (HFM) to solve the linearized scalar equation numerically.
It is found that there is no lower bound on $a$, compared to the case of a negative coupling parameter.
Instead, we find the upper bound on $a$ ($M^2-Q^2\ge a^2$) as the existence condition for the outer horizon.
Importantly, we obtain a 3D graph [$\log_{10}\alpha(a,Q)$] which indicates the onset surface for
spontaneous scalarization of KN black holes clearly in the EMS theory.
In addition, we confirm that the high rotation enhances spontaneous scalarization of
KN black holes in the EMS theory.
Our action of the EMS theory takes the form ~\cite{Herdeiro:2018wub}
\begin{eqnarray}
S_{\rm EMS}=\frac{1}{16\pi}\int d^4x{\sqrt{-g}\Big[R-2\partial_\mu\phi\partial^\mu\phi-f(\phi)F_{\mu\nu}F^{\mu\nu}\Big]}. \label{action}
\end{eqnarray}
Here, the coupling function $f(\phi)$ is important to control a nonminimal coupling of scalar $\phi$
to Maxwell term $F_{\mu\nu}F^{\mu\nu}\equiv F^2$ with $F_{\mu\nu}=\partial_{\mu}A_{\nu}-\partial_{\nu}A_{\mu}$.
We will probe the tachyonic instability of KN black holes by adopting hyperboloidal foliation method.
In this case, we do not need to worry about the sophisticated outer boundary problem \cite{Zenginoglu:2011zz,Thuestad:2017ngu},
since the ingoing (outgoing) boundary condition at the horizon (infinity) to solve  the linearized scalar
equation  will be satisfied automatically.

We organize the present work  as follows. In Sec.~\ref{2s}, we revisit  briefly the linearized scalar
equation in the EMS theory. Then,  we study the spin-charge induced scalarization of KN black holes in Sec.~\ref{3s}.
The Sec.~\ref{4s} is devoted to contributing to conclusions and discussions.

\section{Linearized scalar equation}\label{2s}

The variation of action \eqref{action} with respect to the metric $g_{\mu\nu}$, scalar field $\phi$,
and vector potential $A_\mu$ gives the following  equations:
\begin{eqnarray}
&&R_{\mu\nu}-\frac{1}{2}R g_{\mu\nu}=2\partial _\mu \phi\partial _\nu \phi-(\partial \phi)^2g_{\mu\nu}
+2f(\phi)\Big(F_{\mu\rho}F_{\nu}~^\rho-\frac{F^2}{4}g_{\mu\nu}\Big), \label{g-eql}\\
&&\nabla_{\mu}\nabla^{\mu}\phi -\frac{1}{4}\frac{d f(\phi)}{d\phi}F^2=0, \label{s-equa1}\\
&&\partial_\mu\left(\sqrt{-g}f(\phi)F^{\mu\nu}\right)=0\label{M-eq1}.
\end{eqnarray}
For a coupling function $f(\phi)$, its form  has to  accommodate a GR black hole in the limit
of $\phi\rightarrow0$ (namely, KN black holes  in the Einstein-Maxwell gravity).

To discuss the tachyonic instability of black holes in EMS theory, we consider the linearized
perturbation equation
\begin{eqnarray}
\left(\bar{\square}-\mu_{\rm eff}^2\right)\delta\phi=0, \quad
\mu_{\rm eff}^2=\frac{F^2}{4}\frac{d^2 f}{d\phi^2}(0)\label{per-eq}.
\end{eqnarray}
This instability is characterized by the presence of a
negative mass term ($\mu^2_{\rm eff}<0)$ in the linearized scalar equation.
Until now, many authors adopted different forms of coupling function
$f(\phi)$, such as exponential coupling form $e^{\alpha\phi^2}$~\cite{Herdeiro:2018wub},
quadratic coupling form $1+\alpha\phi^2$~\cite{Myung:2018vug}, hyperbolic cosine coupling
form $\cosh\sqrt{2\alpha}\phi$~\cite{Fernandes:2019rez}, and so on.
Without loss of generality,
we do not choose any precise form of coupling function $f(\phi)$ and only
require that $f(\phi)$ satisfy~\cite{Fernandes:2019rez,Herdeiro:2019yjy}
\begin{eqnarray}\label{fphi}
f(0)=1,\quad \frac{d f}{d\phi}(0)=0, \quad \frac{d^2f}{d\phi^2}(0)=2\alpha. \label{fphi2}
\end{eqnarray}
Therefore, it will make our model more general in the investigation for spontaneous scalarization
of KN black holes in EMS theory.

Without scalar hair, the axisymmetric KN black hole is expressed in terms of the Boyer-Lindquist coordinates as
\begin{eqnarray}
ds^2_{\rm KN} &\equiv& \bar{g}_{\mu\nu}dx^\mu dx^\nu=-\frac{\Delta-a^2\sin^2\theta}{\rho^2}dt^2
-\frac{2a\sin^2\theta(r^2+a^2-\Delta)}{\rho^2}dt d\varphi\nonumber\\
&&+\frac{[(r^2+a^2)^2-\Delta a^2 \sin^2\theta]\sin^2\theta}{\rho^2} d\varphi^2+ \frac{\rho^2}{\Delta}dr^2 +\rho^2 d\theta^2, \label{KN-sol}
\end{eqnarray}
where
\begin{eqnarray}
 \Delta= r^2-2Mr+a^2+Q^2,\quad \rho^2=r^2+a^2\cos^2\theta.\nonumber
\end{eqnarray}
The corresponding vector potential is
\begin{eqnarray}
A=-\frac{Qr}{\rho^2}\left(dt-a\sin^2\theta d\varphi\right).\label{vecp}
\end{eqnarray}
Then, the outer and inner horizons are obtained by imposing  $\Delta=(r-r_+)(r-r_-)=0$ as
\begin{eqnarray}
r_{\pm}=M\pm \sqrt{M^2-a^2-Q^2},
\end{eqnarray}
where one requires the existence condition for the outer horizon ($M^2-Q^2\ge a^2$).
For simplicity, we set the mass of KN black hole to be $M=1$ in the whole article.

We shall compute the tachyonic instability of KN black holes in the EMS theory. Considering Eqs.\eqref{per-eq},\eqref{fphi2}, and \eqref{vecp},
the effective mass squared is obtained as
\begin{eqnarray}
\mu_{\rm eff}^2=-\frac{\alpha Q^2(r^4-6a^2r^2\cos^2\theta+a^4\cos^4\theta)}{\left(r^2+a^2\cos^2\theta\right)^4}.\label{effmass}
\end{eqnarray}
Under the scalar perturbation with $\alpha>0$, the KN black holes may become unstable in the presence of $\mu^2_{\rm eff}<0$.
\begin{figure*}[t!]
   \centering
  \includegraphics{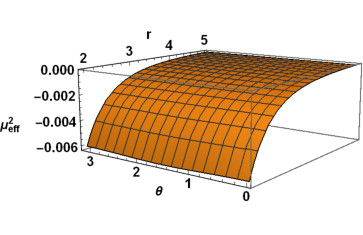}
  \hfill%
  \includegraphics{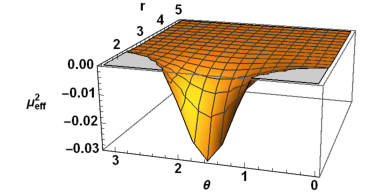}
\caption{ Two 3D graphs for the effective mass term ($\mu^2_{\rm eff}$) with $\alpha=1$, $M=1$, and $Q=0.3$.
These include $r\in[r_+,5]$ and $\theta\in[0,\pi]$. Left: a  graph for $a=0.1$ and $r_+=1.95$ shows whole negative region.
Right: a graph for $a=0.9$ and $r_+=1.32$ represents  negative region around $\theta=\pi/2$
with small positive regions at $\theta=0$ and $\pi$. }\label{fig1}
\end{figure*}

Intuitively, the influence of an effective mass term $\mu^2_{\rm eff}$ on the tachyonic instability
of KN black holes could be understood  from its profiles. From Fig.~\ref{fig1},
one observes that a graph of $\mu^2_{\rm eff}$ with $a=0.1$ shows a whole negative region,
while a graph for $a=0.9$ represents a negative region around $\theta=\pi/2$ [$\mu^2_{\rm eff}=-Q^2/r^4$
for the Reissner-Nordstr\"{o}m (RN) case] and maintains positive near $\theta=0,\pi$.
It appears that the negative region in the $\theta$ direction decreases as $a$ increases.
One observes from  Fig. 1 that for $\alpha=1$, $\mu_{\rm eff}^2(r=r_+,Q=0.3,a=0.9,\theta=\pi/2)=-0.03$
is $20\times \mu_{\rm eff}^2(r=r_+,Q=0.3,a=0.1,\theta=\pi/2)=-0.006$. Therefore, one  conjectures that
the high rotation enhances spontaneous scalarization. In the chargeless limit of $Q\to 0$,
one finds $\mu^2_{\rm eff} \to 0$ (a massless scalar propagation).
 However, its specific form of threshold curve [$\log_{10}\alpha(a,Q)$] will be determined only after performing
numerical computations for a very long time.

\section{Spin-charge induced scalarization }\label{3s}

Before  proceeding, we mention that it is not easy to solve the partial differential equation \eqref{per-eq} directly.
In our previous work~\cite{Lai:2022spn},
we adopted the $(2+1)$-dimensional hyperboloidal foliation method to  solve the linearized scalar equation
numerically with negative coupling parameter. It is worth noting that the time evolution of a linearized scalar
is independent of the sign of coupling parameter $\alpha$. Hence, we adopt this method to explore
the evolution of the linearized scalar with positive coupling parameter.
Using two coordinate transformations and introducing the auxiliary fields
\begin{eqnarray}
\Psi_m = \partial_R u_m, \quad \Pi_m = \partial_T u_m,
\end{eqnarray}
three coupled first-order differential equations are given by~\cite{Lai:2022spn}
\begin{eqnarray}
&&\partial_T u_m = \Pi_m, \label{phi-eq9}\\
&&\partial_{T} \Psi_m=\partial_T\partial_R u_m=\partial_R \Pi_m,\label{phi-eq7}\\
&&\partial_{T} \Pi_m=-(\tilde{B}^{T} \Pi_m+
\tilde{A}^{TR}\partial_{R} \Pi_m +\tilde{A}^{RR}\partial_{R}\Psi_m
+\tilde{A}^{\theta\theta}\partial_{\theta}^2u_m\nonumber\\
&&\qquad\qquad+\tilde{B}^{R}\Psi_m+
\tilde{B}^{\theta}\partial_{\theta} u_m+\tilde{C} u_m). \label{phi-eq8}
\end{eqnarray}
Here  $R$ and $T$ are the compactified radial and
suitable time coordinates in the hyperboloidal foliation method, respectively.
More details of the derivation and the coefficients of the above equations can be
found in Appendix A.

Now the spatial part of  differential equations will be solved by making use of the finite difference method
and  the time evolution can be explored  by using the fourth-order Runge-Kutta integrator.
In particular, the ingoing (outgoing) boundary conditions at the outer horizon (infinity) are satisfied automatically
in $R$ direction on account of the Racz and Toth coordinates. So we do not need to set the ingoing (outgoing) boundary conditions.
For boundary conditions in the $\theta$ direction, the coefficients of Eq.~\eqref{phi-eq8} become
singular at $\theta=0$ and $\theta=\pi$. Therefore, to implement these conditions ~\cite{Lai:2022spn,Gao:2018acg,Pazos-Avalos:2004uyd},
we adopt a staggered grid and add ghost points as
\begin{eqnarray}
&&u_m(T,R,\theta)=u_m(T,R,-\theta),\nonumber\\
&&u_m(T,R,\pi+\theta)=u_m(T,R,\pi-\theta), \quad {\rm for} \quad m=0,\pm 2,\cdots
\end{eqnarray}
and
\begin{eqnarray}
&&u_m(T,R,\theta)=-u_m(T,R,-\theta),\nonumber\\
&&u_m(T,R,\pi+\theta)=-u_m(T,R,\pi-\theta), \quad {\rm for } \quad m=\pmb 1,\pm 3,\cdots.
\end{eqnarray}
During numerical computation, we compute the equation in a domain $(R_+,1)\times(0,\pi)$ with
grids of $201\times 68$ in the $R$ and $\theta$ directions.

\begin{figure}
\centering
\subfigure[$m=0$, initial mode $l=0$ ]{
\includegraphics[width=0.4\textwidth]{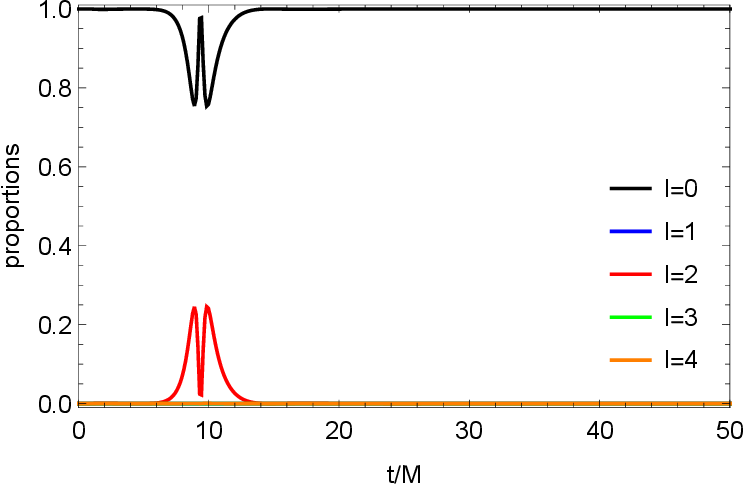}}
\quad
\subfigure[$m=0$, initial mode $l=1$ ]{
\includegraphics[width=0.4\textwidth]{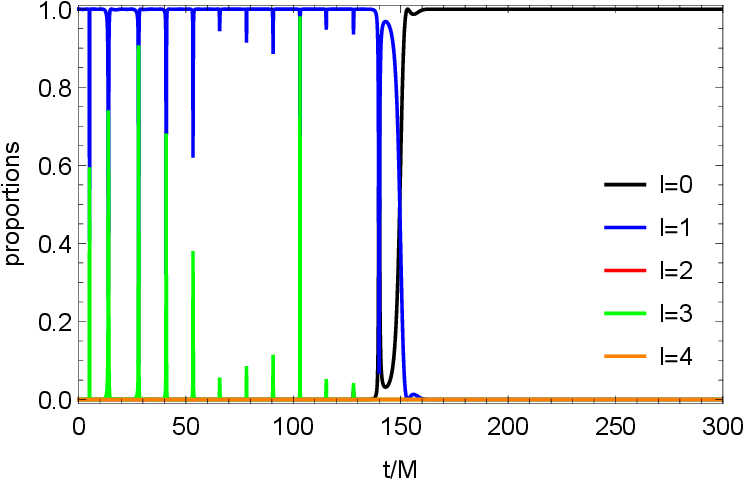}}
\quad
\subfigure[$m=0$, initial mode $l=2$ ]{
\includegraphics[width=0.4\textwidth]{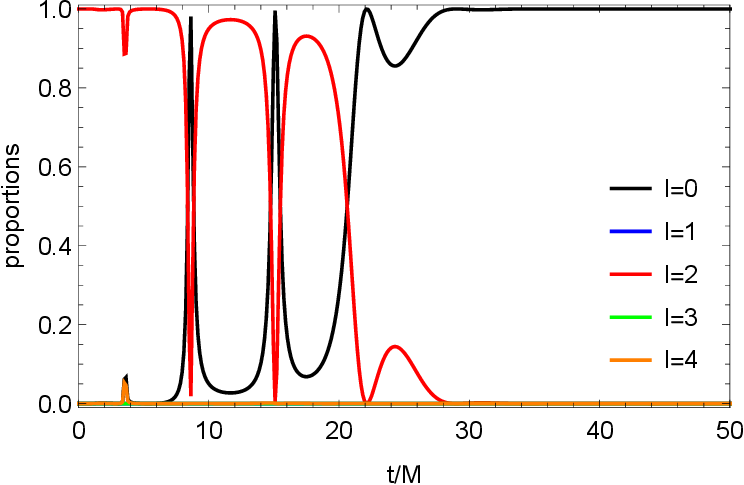}}
\quad
\subfigure[$m=0$, initial mode $l=3$ ]{
\includegraphics[width=0.4\textwidth]{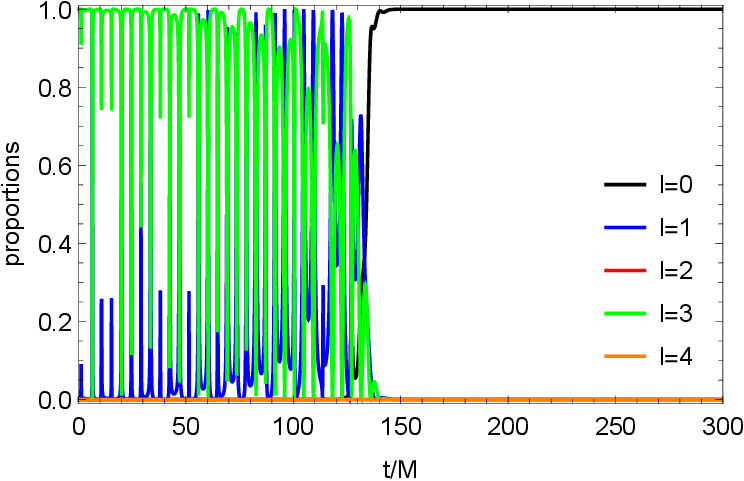}}
\caption{Several modes with $m=0$ and different $l$ numbers are activated by the initial $l$ mode during the process of time evolution.
Here the initial mode is specified to be the one with (a)$l=0$, (b) 1, (c) 2 and (d) 3, respectively. It is found that the
dominant mode is always the $l=m=0$ mode at late enough time, whatever initial mode is chosen.}\label{Fig02}
\end{figure}

As the initial perturbation, we choose a spherically harmonic Gaussian distribution
centered at $R_c$ outside the outer  horizon
\begin{eqnarray}
u_{lm}(T=0,R,\theta)\sim Y_{lm}(\theta,0)e^{-\frac{(R-R_c)^2}{2\sigma^2}}, \label{ini-S}
\end{eqnarray}
where $Y_{lm}(\theta,\phi)$ represent the spherical harmonic functions and $\sigma$ is the width of Gaussian distribution.
In the following, we take $R_c=\frac{R_{+}+1}{2}$ with $R_+=\frac{\sqrt{r_+^2+1}-1}{r_+}$ and $\sigma=\frac{1}{\sqrt{200}}$.
Also, we set $M=1$ so that all quantities are measured in units of $M$.
It points out that although  there is one initial mode with a specified $l$ only,
other $l$ modes with the same index $m$ will be activated during evolving processes.
The $l=m$ mode will have a dominant contribution at late times.
For example, we can evaluate the processes of time evolution for several modes with $m=0$ and
different $l$ numbers, and found that the late-time dominant mode
is always the mode with $l=m=0$, see Fig.~\ref{Fig02}.
Actually, a similar phenomenon has also occurred for Kerr black
holes \cite{Doneva:2020nbb,Gao:2018acg} and KN black holes \cite{Lai:2022spn}.
Hereafter, we consider axisymmetric perturbations with $l=m=0$ only for simplicity.

\begin{figure}[htbp]
\centering
\includegraphics[width=0.45\textwidth]{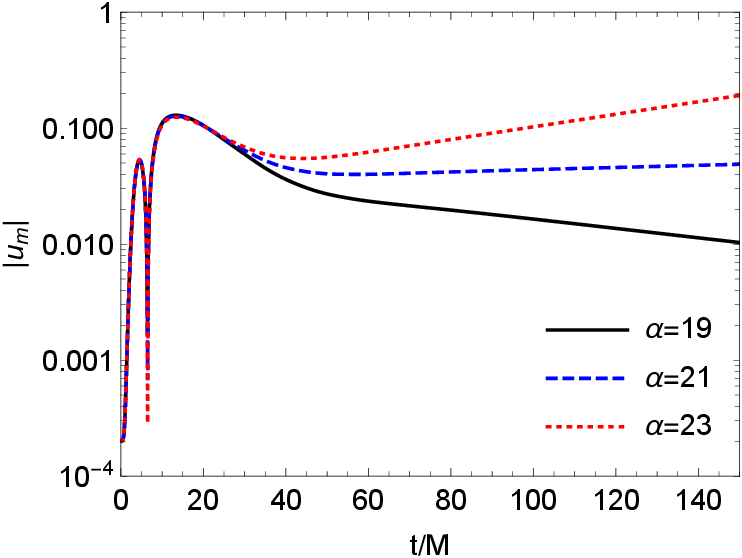}
\caption{Time-domain profiles of a linearized scalar $|u_m|$ for $Q=0.4$ and $a=0.9$ with $M=1$
and different coupling constants $\alpha$.}\label{Fig03}
\end{figure}

Considering different coupling parameter $\alpha$, the time-domain profiles of the $l=m=0$
mode for a linearized scalar  are plotted in Fig.~\ref{Fig03}.
The $\alpha=21$ case represents the threshold (marginal) evolution of instability.

\begin{figure}[htbp]
\centering
\subfigure[]{
\includegraphics[width=0.4\textwidth]{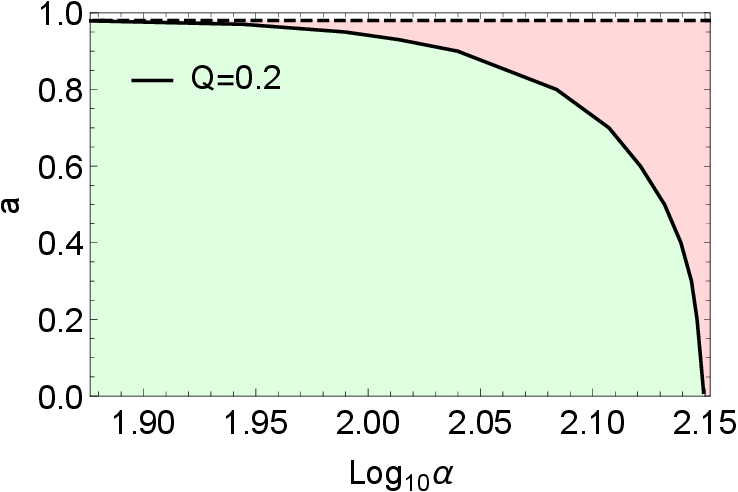}\label{Fig4a}}
\quad
\subfigure[]{
\includegraphics[width=0.4\textwidth]{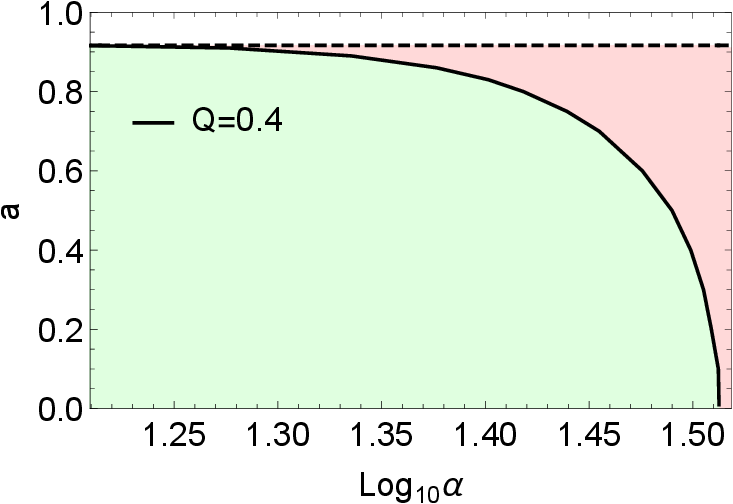}\label{Fig4b}}
\quad
\subfigure[]{
\includegraphics[width=0.4\textwidth]{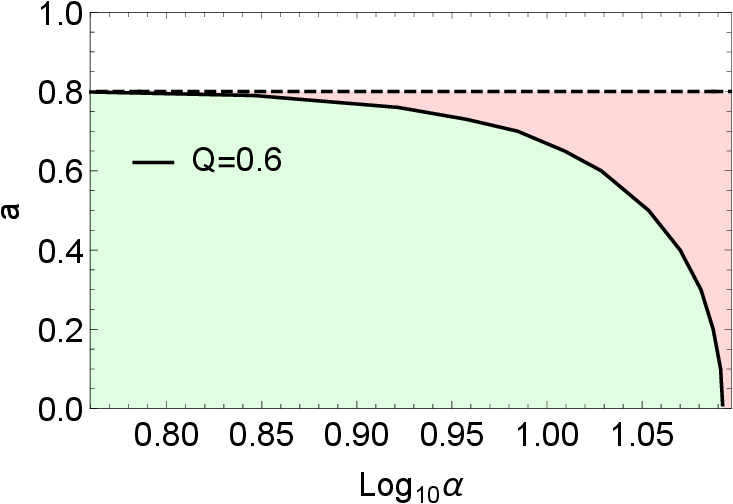}\label{Fig4c}}
\quad
\subfigure[]{
\includegraphics[width=0.4\textwidth]{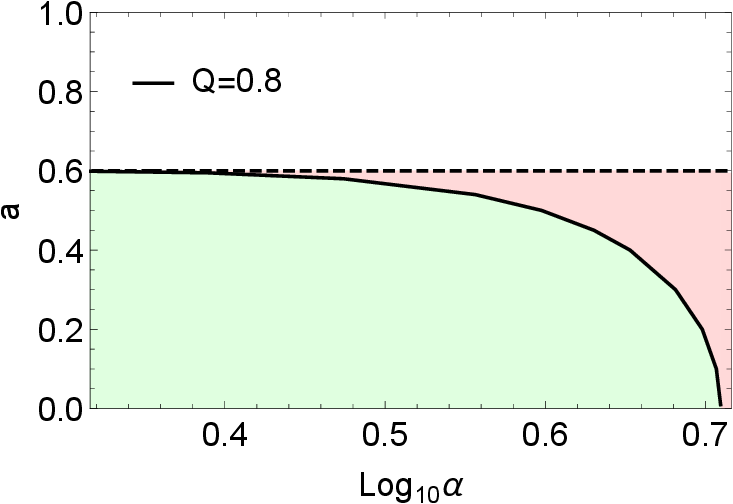}\label{Fig4d}}
\caption{Four threshold curves [$\log_{10}\alpha(a)$] for the dominant $m=0$ mode with  (a)$Q=0.2$ ,(b)0.4, (c)0.6 and (d)0.8, and $M=1$.
The  solid lines denote the boundaries between stable (green) and unstable (red) region. The dashed lines
represent the existence condition ($M^2-Q^2\ge a^2$) of the outer horizon.}\label{Fig04}
\end{figure}

Furthermore, we  make the marginal  curves with different charge $Q$, rotation parameter $a$,
and coupling parameter $\alpha$ for a fixed mode $m=0$.
We find that the tachyonic  instability of KN black holes
happens only in a certain region of the parameter space (see Fig.~\ref{Fig04}).  Choosing $Q=0.2$, for example,
the threshold curve [$\log_{10}\alpha(a)$] starts at $\log_{10}\alpha=\log_{10} 141.041(= 2.149)$ on the $\alpha$ axis,
corresponding to the threshold of unstable RN black holes~\cite{Myung:2018vug}.
It determines the right boundary in the $\log_{10}\alpha(a)$ graph [Fig.~\ref{Fig4a}].
In this case, the instability can be only achieved in the region ($\log_{10}\alpha \ge 2.149$), and this threshold
curve decreases from 2.149 to a smaller value  as $a$ increases from 0 to 0.980 (upper bound) with $Q=0.2$.
Here, the upper bound (denoted by dashed line) is determined by  the existence condition of the outer horizon
($M^2-Q^2\ge a^2$). For $Q=0.2$, 0.4, 0.6 and 0.8, the maximum values (upper bound) of $a$ are  given by  0.980, 0.917, 0.8 and 0.6, respectively.
We note  that the upper bound on  $a$ also decreases as charge  $Q$ increases. Importantly, we find from Fig.~\ref{Fig04} that for all fixed $Q$ and $\alpha$,
there exist unstable scalar  modes for the points in the region (red) above the threshold curve and the KN black holes become unstable in this region.
On the other hand,  for  points in the region (green) below the threshold curve, there are no growing modes and the KN black holes are stable.
\begin{figure}[htbp]
\centering
\includegraphics[width=0.7\textwidth]{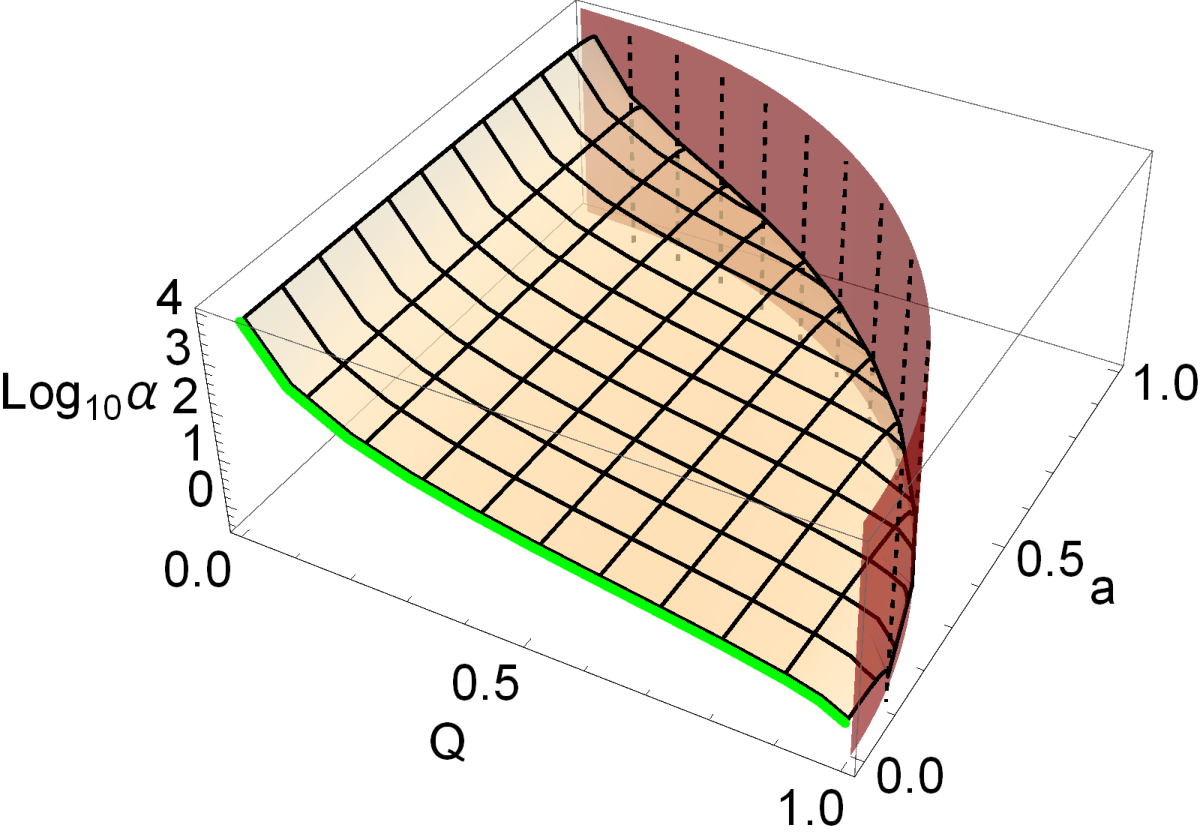}
\caption{The 3D threshold graph [$\log_{10}\alpha(a,Q)$] as function of rotation parameter $a$ and  charge $Q$.
The orange surface denotes the onset surface of  spin-charge induced scalarization for  KN black holes in the EMS theory.
The brown boundary represents the existence condition of outer horizon ($M^2-Q^2\ge a^2$) and the green curve
indicates the threshold curve for the unstable RN black holes. }\label{Fig05}
\end{figure}

In Fig.~\ref{Fig05}, we obtain a 3D graph [$\log_{10}\alpha(a,Q)$] which indicates the onset surface of spin-charge
induced scalarization for KN black holes in the EMS theory. The brown boundary is constructed by combining all dashed
lines shown in Fig.~\ref{Fig04}, which represents the existence condition for the outer horizon ($M^2-Q^2\ge a^2)$.
The orange surface depicts the boundary  between the stable (lower) region and unstable (upper) regions. In the
nonrotating limit of $a\rightarrow 0$, this surface reduces to the green curve, which represents
the threshold curve of tachyonic instability for RN black holes~\cite{Myung:2018vug}.
In the chargeless limit of $Q\to 0(\mu^2_{\rm eff}\to 0)$, one recovers a massless scalar propagation around the Kerr black holes,
which turns out to be stable because there is no room for the unstable region.
This picture indicates the onset of spontaneous scalarization for the KN black holes in the EMS theory clearly.

On the other hand, we also try to adopt the $(2+1)$ time
evolution method to recalculate the late-time tails of
the perturbed scalar field to perform the tachyonic instability
of the KN black holes numerically in the time domain, as shown in Appendix B.
The late-time tails in Fig.~\ref{Fig06} demonstrate that it serves
as an independent check of the previous results.

\begin{figure}[htbp]
\centering
\includegraphics[width=0.4\textwidth]{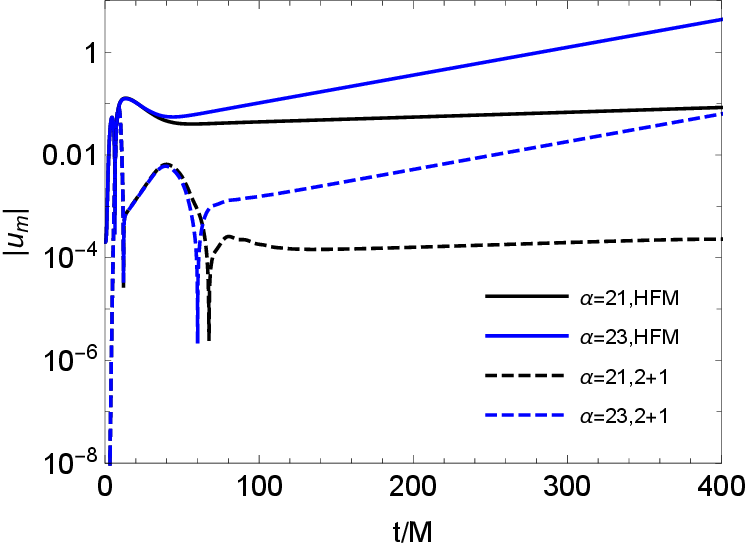}
\caption{The time-domain profiles $(m=0)$ of scalar perturbation with
$Q=0.4$, $M = 1$, $a=0.9$, and different values of $\alpha$ by using two different
numerical methods. This figure implies that the late-time tails of perturbed
scalar fields share the same behaviors.}\label{Fig06}
\end{figure}

\section{Conclusions and discussions}\label{4s}

In this work, we have carried out the tachyonic instability of KN black holes in the EMS theory with
a positive scalar coupling to Maxwell term.  We adopted the hyperboloidal compactification technique to calculate
the $(2+1)$-dimensional  evolution of an initial linearized scalar located outside the outer horizon numerically.
Consequently, we have obtained the 3D  threshold curve [$\log_{10}\alpha(a,Q)$] which describes a
boundary surface between stable (lower region) and unstable (upper region) KN black holes.
Moreover, the high rotation enhances spontaneous scalarization.  Also, the threshold curve decreases rapidly as $a$ increases
and hits the $\alpha$ axis in the nonrotating limit of $a\rightarrow0$. The computation including  nonlinear
effects are expected to quench the tachyonic  instability and lead to constructing  scalarized KN black holes.

Before finishing, we wish to mention  some prospects of this work.
Recently, Doneva \textit{et al}. \cite{Doneva:2020kfv} investigated the spin induced scalarization of Kerr
black hole in the ESGB theory with a massive scalar, and then Zhang~\cite{Zhang:2021btn} has discussed
the Kerr black hole in the dynamical Chern-Simons gravity with a massive scalar. It is  interesting  to generalize
our work  by introducing  a massive scalar. There may  be richer physics involved in studying of the dynamics and the fate of the KN black holes.
In addition, another promising  direction is to construct the scalarized KN black holes in the EMS theory explicitly.

 \vspace{1cm}

{\bf Acknowledgments}

This work is supported by National Key R$\&$D Program of
China (Grant No. 2020YFC2201400).
D. C. Z acknowledges financial support from Outstanding Young Teacher Programme from Yangzhou University, No. 137050368.
M. Y. L acknowledges financial support from the Initial Research Foundation of Jiangxi Normal  University.

 \vspace{1cm}

\newpage

\appendix
\section*{Appendix A}

In this appendix, we will show how to transform the scalar field perturbation equation \eqref{per-eq} into a form
suitable for numerical calculations. This mainly contains two coordinate transformations, which have the advantages
that the time slices are horizon penetrating and connect to future null infinity such that no boundary conditions are needed.

First, we rewrite the KN metric \eqref{KN-sol} in the ingoing Kerr-Schild coordinates $\{\tilde{t},r,\theta,\tilde{\varphi}\}$
by applying the coordinate transformations
\begin{eqnarray}
d\tilde{t}=dt+\frac{2 Mr-Q^2}{\Delta} dr,\quad d\tilde{\varphi}=d\varphi +\frac{a}{\Delta} dr.
\end{eqnarray}
and it becomes
\begin{eqnarray}
 ds^2&=&-\left(1-\frac{2M r-Q^2}{\rho^2}\right)d\tilde{t}^2
 -\frac{2a\left(2M r-Q^2\right)\sin^2\theta}{\rho^2}d\tilde{t} d\tilde{\varphi}+\frac{4M r-2Q^2}{\rho^2}d\tilde{t}dr\nonumber\\
 &&+\left(1+\frac{2M r-Q^2}{\rho^2}\right)d r^2-2a\sin^2\theta\left(1+\frac{2M r-Q^2}{\rho^2}\right) dr d\tilde{\varphi} +\rho^2 d\theta^2\nonumber\\
 &&+\left(r^2+a^2+\frac{(2M r-Q^2)a^2\sin^2\theta}{\rho^2}\right)\sin^2\theta d\tilde{\varphi}^2. \label{KN-KS}
\end{eqnarray}

Given the axial symmetry of the KN geometry, the perturbative variable $\delta\phi$ can be decomposed as
 \begin{eqnarray}
 \delta \phi(\tilde{t},r,\theta,\tilde{\varphi}) =\frac{1}{r} \sum_{m} u_m(\tilde{t},r,\theta) e^{i m \tilde{\varphi}}.
\end{eqnarray}\label{decomposePhi}
Substituting the above ansatz into Eq.~(\ref{per-eq}),
the scalar perturbation equation in the ingoing Kerr-Schild coordinates takes the following form
\begin{eqnarray}
&&A^{\tilde{t}\tilde{t}}\partial_{\tilde{t}}^2u_m+A^{\tilde{t}r}\partial_{\tilde{t}}\partial_r u_m+A^{r r}\partial^2_r u_m\nonumber\\
&&+A^{\theta\theta}\partial_\theta^2u_m+B^{\tilde{t}}\partial_{\tilde{t}}u_m
+B^r\partial_r u_m+B^\theta\partial_\theta u_m+C u_m=0\label{phi-eq4}
\end{eqnarray}
with coefficients
\begin{eqnarray}
&&A^{\tilde{t}\tilde{t}}=\rho^2+2Mr-Q^2,~~A^{\tilde{t}r}=2Q^2-4Mr,~~A^{r r}=-\Delta,~~A^{\theta\theta}=-1,\nonumber\\
&&B^{\tilde{t}}=2M-\frac{2Q^2}{r},~~B^r=\frac{2}{r}(a^2+Q^2-Mr)-2i m a,~~B^\theta=-\cot\theta,  \nonumber\\
&&C=\frac{m^2}{\sin^2\theta}-\frac{2(a^2+Q^2-Mr)}{r^2}+\frac{2i m a}{r}+\mu^2_{\rm eff}\rho^2.\label{coeffs}
\end{eqnarray}

As the second step, it is helpful to introduce the compactified radial coordinate $R$ and the
suitable time coordinate $T$ following Racz and Toth~\cite{Racz:2011qu} with
\begin{eqnarray}\label{RT1}
\tilde{t}=T+h(R), \quad r=R/\Omega(R),
\end{eqnarray}
where the height function $h(R)$ and conformal
factor $\Omega(R)$ are given by
\begin{eqnarray}\label{RT2}
h(R)=\frac{1+R^2}{1-R^2}-4M \ln(1-R^2), \quad \Omega(R)=\frac{1-R^2}{2}.
\end{eqnarray}
As the result of the conformal transformation \eqref{RT1}, Eq.~(\ref{phi-eq4}) becomes
\begin{eqnarray}
&&\partial^2_{T} u_m+\tilde{A}^{T R}\partial_{T} \partial_{R} u_m+\tilde{A}^{RR}\partial^2_R u_m\nonumber\\
&&+\tilde{A}^{\theta\theta}\partial_\theta^2u_m+\tilde{B}^{T}\partial_{T}u_m
+\tilde{B}^R\partial_R u_m+\tilde{B}^\theta\partial_\theta u_m+\tilde{C} u_m=0, \label{phi-eq5}
\end{eqnarray}
where coefficients can be expressed with
\begin{eqnarray}
\{\tilde{A}^{T R},\tilde{A}^{RR},
\tilde{A}^{\theta\theta},\tilde{B}^{T},
\tilde{B}^R,\tilde{B}^\theta,\tilde{C}\}
=\frac{1}{A^{TT}}\{A^{T R},A^{RR},
A^{\theta\theta},B^{T},
B^R,B^\theta,C\}
\end{eqnarray}
and
\begin{eqnarray}
&&A^{TT}=A^{\tilde{t}\tilde{t}}-H A^{\tilde{t} r}+H^2A^{rr},\nonumber\\
&&B^{T}=B^{\tilde{t}}-H B^{r}-\frac{(1-R^2)^2}{2(1+R^2)}H'A^{rr},\nonumber\\
&&A^{RR}=\left(\frac{(1-R^2)^2}{2(1+R^2)}\right)^2A^{rr},\nonumber\\
&&A^{TR}=\frac{(1-R^2)^2}{2(1+R^2)} A^{\tilde{t}r}-\frac{(1-R^2)^2}{1+R^2}HA^{rr},\nonumber\\
&&B^{R}=\frac{(1-R^2)^2}{2(1+R^2)}\left[B^{r}+\left(\frac{(1-R^2)^2}{2(1+R^2)}\right)'A^{rr}\right],
\end{eqnarray}
where the prime denotes the derivative $\frac{d}{dR}$ and $H(R)=\frac{dh}{dr}(R)$.

Finally, by introducing the following auxiliary fields
\begin{eqnarray}
 &&\Psi_m = \partial_R u_m, \\
 &&\Pi_m = \partial_T u_m,\label{phi-eq6}
\end{eqnarray}
one finds the following coupled equations:
\begin{eqnarray}
&&\partial_T u_m = \Pi_m,\\
&&\partial_{T} \Psi_m=\partial_T\partial_R u_m=\partial_R \Pi_m,\\
&&\partial_{T} \Pi_m=-(\tilde{B}^{T} \Pi_m+
\tilde{A}^{TR}\partial_{R} \Pi_m +\tilde{A}^{RR}\partial_{R}\Psi_m
+\tilde{A}^{\theta\theta}\partial_{\theta}^2u_m\nonumber\\
&&\qquad\qquad+\tilde{B}^{R}\Psi_m+
\tilde{B}^{\theta}\partial_{\theta} u_m+\tilde{C} u_m),
\end{eqnarray}
which are first order in space and time.

\section*{Appendix B}

We will calculate the late-time tails of a perturbed scalar field to perform the tachyonic instability
of the KN black holes numerically in the time domain. For the linearized scalar equation \eqref{per-eq},
we introduce a new coordinate $\varphi^*$ and tortoise coordinate $x$ through the transformations~\cite{Zhang:2021btn}
\begin{eqnarray}
  d\varphi^* &=& d\varphi +\frac{a}{\Delta}dr, \nonumber\\
  dx &=& \frac{r^2+a^2}{\Delta}dr. \label{new-coor}
\end{eqnarray}
Then we have a scalar field perturbed equation
\begin{eqnarray}
  &&\left[ (r^2+a^2)^2-
      \Delta a^2\sin^2\theta\right]\partial^2_t \delta\phi
    -(r^2+a^2)^2\partial_x^2\delta\phi
    -2r\Delta\partial_x\delta\phi
   \nonumber\\
    &&+2a\left(2Mr-Q^2\right)\partial_t\partial_{\varphi^*}\delta\phi-2a(r^2+a^2)\partial_x\partial_{\varphi^*}\delta\phi
   -\frac{\Delta}{\sin\theta}\partial_\theta(\sin\theta\partial_\theta\delta\phi)
    \nonumber  \\
    &&-\frac{\Delta}{\sin^2\theta}\partial^2_{\varphi^*}\delta\phi+\Delta\left(r^2+a^2\cos^2\theta\right)\mu_{\rm eff}^2\delta\phi=0,  \label{mscalar-eq}
\end{eqnarray}
where the coordinate $x\in(-\infty,\infty)$ covers the infinite range which  is accessible to an observer
located outside the outer horizon, while one notes the semi-infinite region  $r\in[r_+,\infty)$ when using the radial coordinate $r$.

Taking into account the axial symmetry of \eqref{KN-sol}, the scalar perturbation could be decomposed as
\begin{equation}
\delta\phi(t,x,\theta,\varphi^*)=\sum_{m}\delta\phi(t,x,\theta)e^ {im\varphi^*} \label{s-dec}
\end{equation}
with $m$ as an azimuthal number.
Substituting \eqref{s-dec} into \eqref{mscalar-eq}, we have a (2+1)-dimensional equation
\begin{eqnarray}
  &&\left[ (r^2+a^2)^2-\Delta a^2\sin^2\theta\right]\partial^2_t\delta\phi -(r^2+a^2)^2\partial^2_x\delta\phi
  -\Delta\partial^2_\theta\delta\phi\nonumber\\
  &&+ 2ima{(2Mr-Q^2)}\partial_t\delta\phi-2\left[r\Delta+ima(r^2+a^2)\right]\partial_x\delta\phi-\Delta\cot{\theta}\partial_\theta\delta\phi \nonumber\\
  &&+  \Delta\Big[(r^2+a^2\cos^2\theta)\mu_{\rm eff}^2+\frac{m^2}{\sin^2\theta}\Big]\delta\phi= 0. \label{mscalar-eq2}
\end{eqnarray}
We may  rewrite \eqref{mscalar-eq2} as the (2+1)-dimensional Teukolsky equation
\begin{eqnarray}\label{perturbedEq8-2}
  \partial^2_t\delta\phi + A^{xx}\partial^2_x\delta\phi
  +A^{\theta\theta}\partial^2_\theta\delta\phi  +B^{t}\partial_t\delta\phi + B^{x}\partial_x\delta\phi+ B^{\theta}\partial_\theta\delta\phi +C \delta\phi= 0
\end{eqnarray}
whose coefficients take the forms
\begin{eqnarray}\label{coeff8-0}
  A^{tt} &=& \left[ (r^2+a^2)^2-\Delta a^2\sin^2\theta\right] , \nonumber\\
  A^{xx} &=&-\frac{(r^2+a^2)^2}{A^{tt}},
  \nonumber\\
  A^{\theta\theta} &=&-\frac{\Delta}{A^{tt}},  \nonumber\\
  B^{t} &=& \frac{ 2ima{(2Mr-Q^2)}}{A^{tt}},
  \nonumber\\
  B^{x} &=& -\frac{ 2r\Delta+2ima(r^2+a^2)}{A^{tt}},
  \nonumber\\
  B^{\theta} &=&-\frac{ \Delta\cot{\theta}}{A^{tt}},
  \nonumber\\
  C &=& \frac{\Delta}{A^{tt}}\left[(r^2+a^2\cos^2\theta)\mu_{eff}^2+\frac{m^2}{\sin^2\theta}\right].
\end{eqnarray}
 We note that for $Q=0$, Eq.~\eqref{perturbedEq8-2} reduces exactly to Eq. (12) in Ref.~\cite{Zhang:2021btn}.
At this stage, we introduce the three auxiliary fields defined by
\begin{eqnarray}
  \Phi &\equiv& \delta\phi, \nonumber\\
  \Psi &\equiv& \partial_x \Phi, \nonumber\\
  \Pi  &\equiv& \partial_t \Phi. \label{aux-f}
\end{eqnarray}
Then, Eq.~\eqref{perturbedEq8-2} can be rewritten as
\begin{eqnarray}\label{perturbedEq8-3}
  \partial_t\Pi
  = -\left(  A^{xx}\partial_x\Psi
  +A^{\theta\theta}\partial^2_\theta\Phi
  +B^t\Pi + B^x\Psi +B^\theta\partial_\theta\Phi+C\Phi\right).
\end{eqnarray}

Dividing the fields into real and imaginary parts
\begin{eqnarray}
% \nonumber % Remove numbering (before each equation)
  \Phi =\Phi_R + i\Phi_I,\quad
  \Psi= \Psi_R + i\Psi_I,\quad
  \Pi = \Pi_R + i\Pi_I,
\end{eqnarray} \label{ri-sep}
Eq.~\eqref{perturbedEq8-3} is separated into two equations
\begin{eqnarray}
 \partial_t\Pi_R=&-&\Big( A^{xx}\partial_x\Psi_R
  +A^{\theta\theta}\partial^2_\theta\Phi_R \nonumber \\
&-&B^t_I\Pi_I + B^x_R\Psi_R- B^x_I\Psi_I+B^\theta\partial_\theta\Phi_R+C\Phi_R\Big),
  \label{eqsv-1}\\
 \partial_t\Pi_I=&-&\Big(  A^{xx}\partial_x\Psi_I
  +A^{\theta\theta}\partial^2_\theta\Phi_I \nonumber \\
&+&B^t_I\Pi_R + B^x_I\Psi_R+ B^x_R\Psi_I +B^\theta\partial_\theta\Phi_I+C\Phi_I\Big).\label{eqsv-2}
\end{eqnarray}
Introducing  $u= (\Phi_R,\Phi_I,\Psi_R,\Psi_I,\Pi_R,\Pi_I)^T$, these equations can be rewritten  compactly as
\begin{eqnarray}
  \partial_t u &=& (G\partial_x+Y)u, \label{comp-eq}
\end{eqnarray}
where
\begin{eqnarray}\label{perturbedEq3}
  G &=&  \left(
           \begin{array}{cccccc}
             0 & 0 & 0 & 0 & 0 & 0 \\
             0  & 0 & 0 & 0 & 0 & 0 \\
             0 & 0 & 0 & 0 & 1 & 0 \\
             0 & 0 & 0 & 0 & 0 & 1 \\
             0 & 0 & G_{53} & 0 & 0 & 0 \\
             0 & 0 & 0 & G_{64} & 0 & 0 \\
           \end{array}
         \right),
  \label{G-mat}\\
   Y &=&  \left(
           \begin{array}{cccccc}
             0 & 0 & 0 & 0 & 1 & 0 \\
             0  & 0 & 0 & 0 & 0 & 1 \\
             0 & 0 & 0 & 0 & 0 & 0 \\
             0 & 0 & 0 & 0 & 0 & 0 \\
             Y_{51} & 0 & Y_{53} & Y_{54} & 0 & Y_{56} \\
             0 & Y_{62}  & Y_{63}  & Y_{64}  & Y_{56}  & 0 \\
           \end{array}
         \right) \label{Y-mat}
\end{eqnarray}
with matrix elements
\begin{eqnarray}
  G_{53} &=& G_{64}=-A^{xx}, \nonumber \\
  Y_{51} &=& Y_{62}= -(A^{\theta\theta}\partial_\theta^2+B^\theta\partial_\theta+C),
  \nonumber\\
  Y_{53} &=& Y_{64}=B^x_R, \nonumber\\
  Y_{54} &=& -Y_{63}=-B^x_I,
  \nonumber\\
  Y_{56}&=&Y_{65}=B^t_I. \label{mat-coeff}
\end{eqnarray}
The derivatives in $x$ and $\theta$ directions are approximated by making use of a finite difference method,
while the time evolution is carried out by adopting the fourth-order Runge-Kutta integrator.
We introduce the boundary conditions: ingoing waves at the outer horizon ($x=-\infty$) and outgoing waves at infinity ($x=\infty$).
At the poles of $\theta=0,\pi$, one imposes the boundary condition of $\Phi|_{\theta=0,\pi}=0$   for $m\not=0$, whereas $\partial \Phi|_{\theta=0,\pi}=0$   for $m=0$.

\end{document}